\documentclass[aip,amsmath,amssymb,preprint]{revtex4-1}

\usepackage{graphics}
\usepackage{graphicx}
\usepackage{color}
\usepackage{amssymb}
\usepackage{float}
\usepackage[utf8]{inputenc}
\usepackage[T1]{fontenc}
\usepackage{mathptmx}

\draft 

\begin{document}


\title{Photoexcited organic molecules $en~route$ to highly efficient autoionization} 



\author{Sesha Vempati}
\altaffiliation{Indian Institute of Technology Bhilai}
\affiliation{Fritz Haber Institute of the Max Planck Society, Department of Physical Chemistry, 
Faradayweg 4-6, 14195 Berlin, Germany}

\author{Lea Bogner}

\author{Clemens Richter}

\author{Jan-Christoph Deinert}
\altaffiliation{Helmholtz-Zentrum Dresden-Rossendorf}

\author{Laura Foglia}
\altaffiliation{Elettra-Sincrotrone Trieste S.C.p.A. }

\author{Lukas Gierster}

\author{Julia St\"ahler}


\date{\today}

\begin{abstract}
The conversion of optical and electrical energy in novel materials is key to modern optoelectronic and light-harvesting applications. Here, we investigate the equilibration dynamics of photoexcited 2,7-bis(biphenyl-4-yl)-2',7'-ditertbutyl-9,9'-spirobifluorene (SP6) molecules adsorbed on ZnO(10-10) using femtosecond time-resolved two-photon photoelectron (2PPE) and optical spectroscopy. We find that, after initial ultrafast relaxation on fs and ps timescales, an optically dark state is populated, likely the SP6 triplet (T) state, that undergoes Dexter-type energy transfer ($r_\mathrm{Dex}=1.3$~nm) and exhibits a long decay time of 0.1~s. Because of this long lifetime a photostationary state with average T-T distances below 2~nm is established at excitation densities in the $10^{20}\mathrm{cm}^{-2}~\mathrm{s}^{-1}$ range. This large density enables decay by T-T annihilation (TTA) mediating autoionization despite an extremely low TTA rate of $k_{\mathrm{TTA}}=4.5\cdot10^{-26}$~m$^3$s$^{-1}$. The large external quantum efficiency of the autoionization process (up to 15~$\%$) and photocurrent densities in the mA~cm$^{-2}$ range offer great potential for light-harvesting applications.

\end{abstract}

\pacs{}

\maketitle 

\section{Introduction}
\label{intro}
Electronic and excitonic coupling between the excited states of molecules and across hybrid interfaces \cite{blumstengel2008}
are vital aspects for various applications such as photovoltaics\cite{Cheng2016, Goldschmidt2015}, photochemical reactions\cite{Borjesson2013}, solar energy conversion\cite{Wondraczek2015}, and opto-electronics\cite{Schlesinger2015}. The strength and nature of coupling between any two excited states as well as the energy level alignment at an interface are dependent on the molecular architecture, the degree of molecular order, and bonds as well as molecular arrangements at interfaces\cite{Friedlein2009,Parker1962,Pope1966,Haarer1971}. One overarching goal of all these applications is the conversion of energy while minimizing energy losses. Particularly when aiming at the exploitation of solar radiation, this is a major challenge, as the band gap of the absorbing species sets a lower boundary for the convertible photon energies while high energy photons are absorbed, but a significant portion of their energy is dissipated into heat\cite{Shockley1961}. One approach to maximize the number of charge carriers generated by light is to use singlet fission, where a high energy singlet state decays to two triplet states, thereby duplicating the number of charge carriers that can be harvested\cite{Tritsch2013}. In the inverse process, triplet-triplet annihilation (TTA) through photon upconversion, two triplets interact by an electron-exchange Dexter mechanism that populates a singlet exciton at higher energy. Recombination leads to emission of an upconverted photon\cite{Cheng2016,Goldschmidt2015,Borjesson2013,Wondraczek2015,Parker1962,Parker1963, Singh2009,Gray2014,Arnold1979}, which can, for instance, drive photochemical reactions 
or, in the case of incident infrared radiation, to the generation of visible photons that can be harvested by conventional solar cells\cite{Gray2014}.

Whether or not TTA occurs in organic compounds depends on many aspects. For instance, anthracene solutions exhibit photon upconversion\cite{Parker1962} whereas the crystalline counterpart shows 
autoionization\cite{Pope1966,Haarer1971}. The latter can be induced by various excited state coupling schemes: Singlet/singlet exciton (singlet fusion)\cite{Wessel1990,Ono1998}, free electron/charge-transfer exciton (CTX)\cite{Arnold1979}, CTX-CTX annihilation\cite{Pope1966} or free electron/singlet exciton\cite{Haarer1971}, as observed in naphthalene trimer\cite{Wessel1990}, perylene films\cite{Friedlein2009} and fluoranthene crystals \cite{Ono1998}. 
In these cases, the higher excited singlet is above the vacuum level, while, notably, the excitation energies are lower than the first ionization potential of the molecules. Remarkably, to our knowledge, autoionization mediated by TTA has not been observed to date. 

For both photon upconversion and autoionization, the cooperative interaction of two excited states provides the required energy while their lifetimes, densities, and mobilities determine if the promotion of the electron to the higher excited state is possible. Energy transfer in organic compounds usually occurs by either a F\"orster or a Dexter mechanism, while the former process mostly occurs in the case of singlet and the latter in the case of triplet excitations. The F\"orster transfer, which occurs \emph{via} dipole-dipole interaction, shows a rate constant\cite{Narayan}

\begin{equation}
   k_{\mathrm{F}}(r)=k_{\mathrm{D}}\left(\frac{r_{\mathrm{F}}}{d}\right)^6,
    \label{eq:kF}
\end{equation}

\noindent which is determined by the distance of donor and acceptor $d$, the excited state lifetime $\tau_{\mathrm{D}}=(k_{\mathrm{D}}^{-1})$, and the F\"orster radius $r_{\mathrm{F}}$, at which 50$~\%$ of the excitations transfer to the accepting molecule. In the case of Dexter transfer, the involved carriers are promoted to the acceptor by two charge transfer processes. These require wave function overlap, which is why Dexter transfer is a short-range process, usually occuring across few nanometer distances. The rate constant can be written as\cite{Narayan}

\begin{equation}
  k_{\mathrm{Dex}}(r)=k_{\mathrm{D}}\cdot\mathrm{exp}\left(\frac{2r_{\mathrm{Dex}}}{L}\left(1-\frac{d}{r_{\mathrm{Dex}}}\right)\right),
    \label{eq:kDex}
\end{equation}

\noindent where $L$ is van-der-Waals distance of the molecules and $r_{\mathrm{Dex}}$, analogous to F\"orster transfer, the distance at which 50$~\%$ of the excitations transfer to the accepting molecule. This results from the quantum yield for energy transfer:

\begin{equation}
   \Phi_{\mathrm{ET}}(r)=\frac{k_{\mathrm{ET}}(r)}{k_{\mathrm{ET}}(r)+k_{\mathrm{D}}}.
    \label{eq:QY}
\end{equation}

Eqs.~\ref{eq:kF} and \ref{eq:kDex} illustrate the importance of the lifetime $\tau_{\mathrm{D}}$ of the excited state. It can be reduced when the molecules are adsorbed on an inorganic substrate, as interfacial charge separation competes with the intrinsic dynamics of the excited molecules\cite{Blumstengel2008b,Schlesinger2015,Foglia2016}. One example is our earlier work on 2,7-bis(biphenyl-4-yl)-2',7'-ditertbutyl-9,9'-spirobifluorene (SP6) molecules grown on a ZnO(10-10) substrate\cite{Foglia2016}. SP6 grows with $\rho=1.14\cdot 10^{27}\mathrm{m}^{-3}$ in quite dense amorphous films\footnote{The SP6 density was determined by Mino Sparenberg, Humboldt University Berlin.}
, its absorption sets in at 3.2~eV and is maximal at 3.6~eV\cite{Blumstengel2008b}. Femtosecond time-resolved optical spectroscopy combined with photoluminescence spectroscopy unveiled that weak electronic coupling between singlets localized on the sexiphenyl (6P) and the spiro-linked  biphenyl (2P) moiety of SP6 leads to two independent relaxation pathways as illustrated by the inset in Fig. 1(a)\cite{Foglia2016}.  The recombination of the exciton localized on the sexiphenyl unit (X$_{6\mathrm{P}}$, blue) leads to photoluminescence (PL) which is limited by charge separation at the ZnO interface. This PL exhibits a vibrational progression that is likely resulting from a coupling to the CC-stretch and CH-bend modes of the central fluorenes of SP6\cite{Staehler2013}. In addition, photoexcitation also leads to the population of the biphenyl unit (X$_{2\mathrm{P}}$, green) that does not decay by light emission, but serves as an intermediate state in the formation process (230 ps) of a dark state (DS) with a lifetime exceeding 5~$\mu$s. Unfortunately, neither the origin, energetic position, nor the coupling of this DS to other states could be accessed in these previous experiments\cite{Foglia2016}.

In the present study, we unfold the dynamics in SP6 molecules following optical excitation using tr-2PPE spectroscopy. Remarkably, the evolution of the femtosecond time-resolved photoelectron intensity coincides with the previously observed ultrafast dynamics in the optical experiments. Moreover, we show that the 2PPE spectra are dominated by a spectral signature that originates from autoionization of the SP6 molecules, which is a consequence of the cooperative interaction of two SP6 molecules in the DS, which is likely the triplet state. A simple rate equation model reproduces the non-linear excitation density dependence of the electron emission intensity and supports autoionization of SP6 molecules through TTA. We show that the DS decays with a large time constant of 0.1~s and determine the rate constant for TTA annihilation $k_{\mathrm{TTA}}=4.5\cdot10^{-26}$~m$^3$~s$^{-1}$. The TTA is mediated by Dexter energy transfer with $r_\mathrm{Dex}=1.3$~nm that is enabled by the long lifetime of the excitation and the large density of the SP6 molecules. The resulting electron emission has a large external quantum efficiency (EQE) of up to 15$\%$ and a photocurrent density of 1.5~mA/cm$^2$ at solar UV photon fluxes. Considering that these competitive values result from a mono-molecular organic film, electron emission through TTA might be a promising route for future light harvesting devices.

\begin{figure*}
\includegraphics[width=1.0\columnwidth]{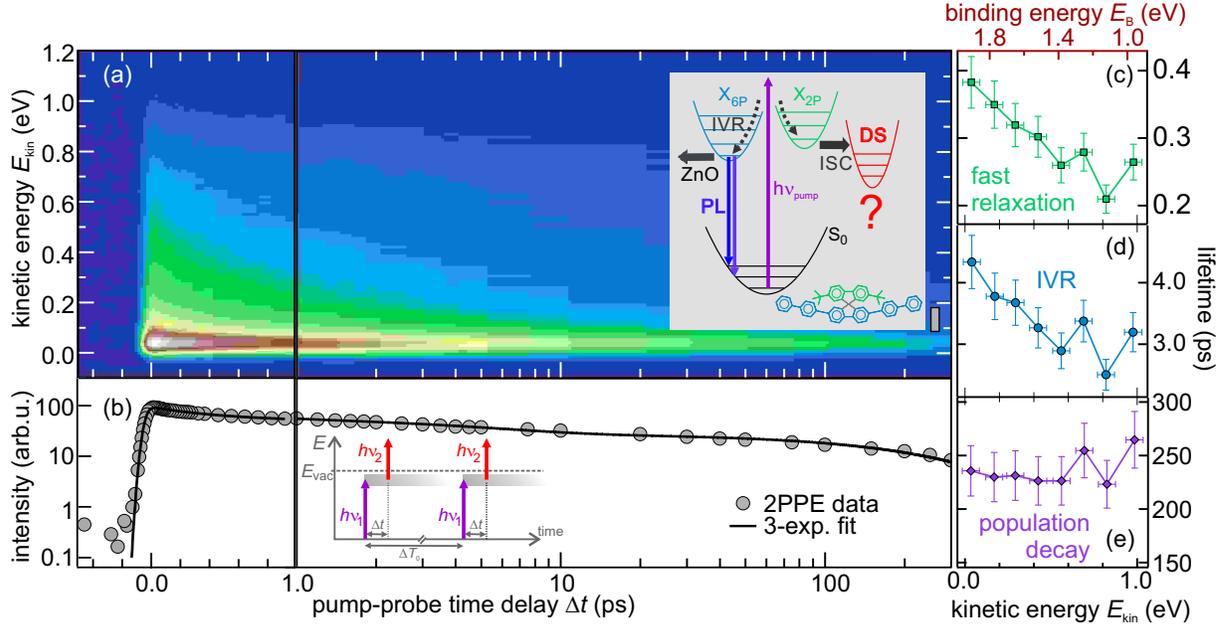}%
\caption{(a) Photoinduced photoelectron intensity (false colors) after subtraction of the constant single color background of 21 nm SP6/ZnO(10-10). The data is plotted as a function of pump-probe pulse time delay (horizontal axis) and kinetic energy (vertical axis) measured using h$\nu_{pump}$ = 3.86 eV and h$\nu_{probe}$= 1.93 eV. Inset: elementary processes following optical excitation of SP6 molecules: internal vibrational relaxation (IVR), photoluminescence (PL) through electron-hole recombination of excitons in the 6P unit of the molecules (X$_{\mathrm{6P}}$) competing with charge separation at the SP6/ZnO interface, intersystem crossing (ISC) to the dark state. (b) Exemplary, energy-integrated ($E_{\mathrm{kin}}=0.11 - 0.23$~eV) cross correlation trace (electron intensity as a function of time delay, grey markers) and triple-exponential fit (black). The inset shows the pump-probe scheme at a pump-probe time delay $\Delta t$and inverse repetition rate $\Delta \mathrm{T}$. Panels (c) – (e) depict the energy dependent fit parameters.\label{2PPE}}%
\end{figure*}

\section{Experimental Details}
\label{Exp}
The samples are prepared in an ultrahigh vacuum (UHV) chamber operating at a base pressure of ~10$^{-10}$~mbar. The ZnO(10-10) surface (MaTecK GmbH) is cleaned by repeated cycles of Ar$+$ sputtering (10 min, $p_{\mathrm{Ar}} = 2.0 \cdot 10^{-6}$~mbar, 750~eV at 300~K) followed by 30 min annealing at 950~K (ramping rate of 30~K/min). The SP6 molecules are purchased from Merck are thoroughly outgassed and evaporated at 580 K from a Knudsen cell onto a freshly prepared surface (at 300~K) to form amorphous  films with nominal thicknesses of up to 25~nm. A pre-calibrated quartz crystal microbalance is employed to determine the thickness of the film. For 2PPE spectroscopy, these samples are  transferred \emph{in situ} into the neighbouring analysis chamber (base pressure 10$^{-11}$~mbar), for optical spectroscopy they are moved through air to an optical cryostat with a base pressure of $10^{-6}$~mbar. In all the reported measurements the sample was kept at 100~K using liquid nitrogen.

Static and time-resolved 2PPE experiments are performed using a regeneratively amplified femtosecond (40 fs pulses) laser system (Ti:Sa, Coherent RegA, 40-200 kHz) which provides a fundamental photon energy of 1.55 eV. The repetition rate (1/$\Delta T_{\mathrm{0}}$) of the laser determines the time-duration $\Delta T_{\mathrm{0}}$ between two subsequent pulses h$\nu_1$, see inset of Fig.~\ref{2PPE}b. A fraction of the total power is directed to an optical parametric amplifier which generates photon energies from 1.85 to 2.55~eV. 3.10, 4.65 and 6.2~eV photons are obtained by frequency doubling, tripling and quadrupling the 1.55~eV energy photons, respectively, while 3.45, 3.62, 3.85 and 4.11~eV photons are obtained by frequency doubling of the output of the optical parametric amplifier. The pulse duration is evaluated from the high energy photoemission cross correlation signal of two laser pulses on a clean tantalum foil inside the UHV chamber. The kinetic energy ($E_{\mathrm{kin}}$) and intensity of the emitted electrons are detected using a hemispherical electron analyzer (PHOIBOS 100, SPECS GmbH) and referenced to the Fermi energy ($E_{\mathrm{F}}$) of the tantalum sample holder which is in electrical contact with the sample surface. A bias voltage of –1.0 V with respect to the analyzer is applied to the sample enabling the detection of electrons of zero $E_{\mathrm{kin}}$. As discussed in detail previously\cite{Staehler2017}, the low energy secondary electron cut-off ($E_{\mathrm{S}}$) in the photoelectron spectra provides the work function ($\Phi$) of the surface when the spectra are plotted as a function of final state energy axis with respect to $E_{\mathrm{F}}$. Its width is an upper limit for the energy resolution of 100 meV.\cite{Staehler2017}. In femtosecond time-resolved 2PPE, a first laser pulse h$\nu_1$ photoexcites the sample and, at a variable time delay $\Delta t$, a second laser pulse $h\nu_2$ photoemits the remaining excited electron population as illustrated by the inset of Fig.~\ref{2PPE}b.

The same laser system was used for the transient transmission experiments. As described earlier in detail\cite{Foglia2016}, 800~nm pulses were used to generate a white light continuum (WLC, 1.77 - 2.4~eV) by focussing into a sapphire crystal and temporally compressed to 20~fs using a deformable mirror as described in Ref.\cite{Wegkamp2011}. In a similar pump-probe scheme as in 2PPE, h$\nu_1$ photoexcites the sample and the resulting changes to the transmitted white light are probed after $\Delta t$. 10 nm band width color filters were used to select the desired h$\nu_2$ and the optical signal detected using a photodiode and a lock-in amplifier. Transient transmission values are normalized to the ground state transmission signal.

\section{Results and Discussion}

\subsection{Ultrafast Dynamics in Photoexcited SP6}
\label{Ultrafast}
We begin by examining the non-equilibrium dynamics of photoexcited SP6 on ultrafast timescales using time-resolved 2PPE. Fig.~\ref{2PPE}a shows an exemplary measurement of the ultrafast response of a 21~nm SP6 film on ZnO(10-10). The photoinduced change of the 2PPE intensity is plotted in false colors and as a function of the electron kinetic energy $E_{\mathrm{kin}}$ (vertical axis) and the time delay between pump and probe laser pulses (horizontal axis, logarithmic for $\Delta t>1$~ps). From $\Delta t=0$ on, a spectrally broad signal is observed that decays on long ps timescales. We extract the time constants of this decay by integrating the 2PPE intensity in eight contiguous 100~meV intervals (see exemplary pump-probe cross correlation trace in Fig.~\ref{2PPE}b, markers). The data can be reproduced well by fitting triple-exponential decay functions, convolved with the Gaussian temporal laser pulses' envelope, to these traces (black curve in Fig.~\ref{2PPE}b). Fig.~\ref{2PPE}c-e depict the energy-dependent relaxation times for the fast (green), intermediate (blue), and slow (purple) time constants. 

\begin{figure}
\includegraphics[width=0.5\columnwidth]{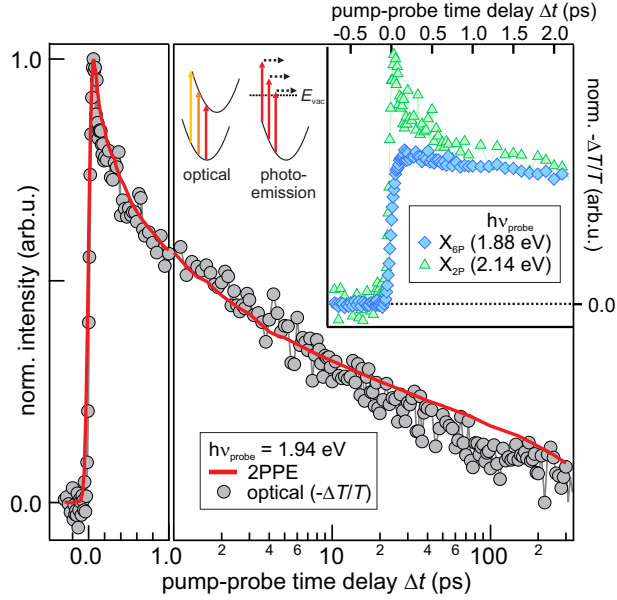}%
\caption{Comparison of time-dependent 2PPE and optical response for the same probe photon energy and similar SP6 films of 21 and 20 nm thickness, respectively. Optical data was originally published by Elsevier in Ref.\cite{Foglia2016}. h$\nu_{\mathrm{pump}}$ = 3.86 and 3.70~eV were used in the 2PPE and the optical experiment, respectively. The techniques monitor the excited state's properties differently, on relative and absolute energy scales, as illustrated by the diagram. Inset: Only the optical response of the X$_{\mathrm{2P}}$ resonance exhibits dynamics on sub-ps timescales.\label{2PPE_Opt}}%
\end{figure}

Before discussing these parameters quantitatively, we compare the time-dependence of the 2PPE signal to the time-resolved \emph{optical} response of a similar SP6 film. Such comparison is not straightforward, because the optical spectroscopy is not performed \emph{in situ} under UHV conditions and as the two techniques probe the non-equilibrium properties of a sample differently as illustrated by the energy level diagrams in Fig.~\ref{2PPE_Opt}. While optical spectroscopy is sensitive to energy resonances in a given system, photoelectron spectroscopy provides the absolute binding energies of electronic levels without the need of a resonance condition. However, in molecular systems, the energy of electronic states is highly sensitive to molecular rearrangements and distortions. The resulting inhomogeneous spread of energy levels can have a much stronger impact on absolute energy scales as probed by 2PPE than on the relative level alignment within the individual molecules of a film as measured by optical spectroscopy\cite{Foglia2016}. In order to still compare the 2PPE and optical data, we integrate the 2PPE intensity across the whole energy range of the spectrally broad signal shown in Fig.~\ref{2PPE}a and compare it (red curve) in Fig.~\ref{2PPE_Opt} to the change of transmission $\Delta T/T$ for the same probe photon energy h$\nu_{\mathrm{probe}}=1.94$~eV (markers). Note that $\Delta T/T<0$ due to excited state absorption and has been multiplied by $-1$. The two data sets are, up to ca. 20~ps, in nearly perfect agreement. For larger pump-probe delays, the 2PPE signal shows a slightly slower decay. 

The coincidence of the non-equilibrium response of the two complementary techniques allows for the following assignment of the different time constants extracted from the 2PPE data in Fig.~\ref{2PPE}c-e:

\begin{itemize}

\item[c)] The fast time constants vary from 200 to 400~fs as a function of kinetic energy and, as such, as a function of vertical binding energy $E_{\mathrm{B}}= h\nu_{\mathrm{probe}}-E_{\mathrm{kin}}$ (top red axis), reflecting that electrons with smaller binding energies have a larger phase space for scattering events and, thus, a shorter lifetime. Due to the low excitation densities in our experiment, we exclude electron-electron scattering as a decay mechanism and assign this time constant to fast relaxation due to coupling to the many high-energy vibrational modes of SP6 \cite{Staehler2013} as, for instance, the CC-stretch modes mentioned in the introduction.
The inset of Fig.~\ref{2PPE_Opt} compares the transient \emph{optical} signal of the two excited state resonances X$_{\mathrm{6P}}$ (diamonds) and X$_{\mathrm{2P}}$ (triangles). Clearly, the fast dynamics occur in the X$_{\mathrm{2P}}$ resonance only. We conclude that the fast relaxation occurs in X$_{\mathrm{2P}}$, which seems to be strongly coupled to high-energy vibrations that allow for fast relaxation.

\item[d)] As for the fast time constants, also the intermediate ones show an energy dependence ranging from 3 to 4~ps for increasing vertical binding energies (Fig.~\ref{2PPE}d, top axis). These lie well within the range of internal vibrational relaxation in X$_{\mathrm{6P}}$ and X$_{\mathrm{2P}}$, $\tau_{\mathrm{IVR}}=$2 to 6 ps as determined in our previous work\cite{Foglia2016}. 

\item[e)] The slow relaxation time is, within error bars, constant for all $E_{\mathrm{kin}}$ and $E_{\mathrm{B}}$, respectively. This time constant is with 240(30)~ps in good agreement with the population decay of X$_{\mathrm{2P}}$ \emph{via} intersystem crossing that occurs in SP6 in $\tau_{\mathrm{ISC}}=230(50)$~ps as determined previously \cite{Foglia2016}. 
\end{itemize}

Based on these results, we conclude that time-resolved 2PPE of photoexcited SP6 exhibits a nearly one-to-one correspondence to previous optical experiments on a similar sample. The 2PPE response is dominated by the relaxation dynamics in X$_{\mathrm{2P}}$, the initial state for ISC. Unfortunately, inhomogeneous broadening on absolute energy scales inhibits spectral characterization of the X$_{\mathrm{6P}}$ and the X$_{\mathrm{2P}}$ state by 2PPE spectroscopy. This is different in the case of the dark state that occurs as a time-\emph{in}dependent background in all 2PPE spectra as discussed in the following section.

\subsection{Microsecond Dynamics of a Photostationary State Probed by Autoionization}
\label{AI}

\begin{figure}
\includegraphics[width=0.5\columnwidth]{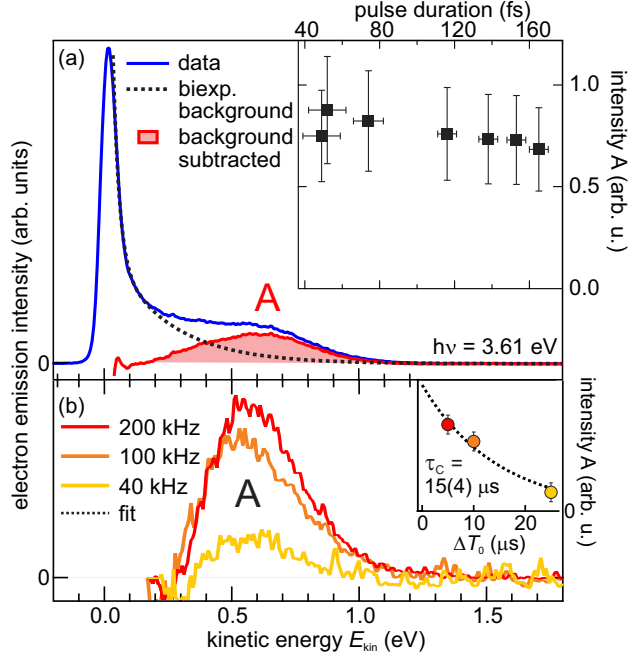}%
\caption{Single-color 2PPE spectra of SP6/ZnO(10-10). (a) Illustration of secondary electron background (dotted curve) subtraction, h$\nu=3.61$~eV, inset: The intensity of peak A is independent of the pulse duration, h$\nu=3.1$~eV. (b) Peak A for varying laser repetition rates, inset: Intensity of A decreases exponentially with $\Delta T_{\mathrm{0}}$, h$\nu=3.6$~eV.\label{lifetime}}%
\end{figure}

Fig.~\ref{lifetime}a shows a one-color 2PPE spectrum (blue) measured with $h\nu=3.61$~eV and with a repetition rate of $\Delta T_0^{-1}=200$~kHz. It is dominated by an intense secondary electron signal which is cut-off at low energies due to the ionization threshold (work function $\Phi=3.3$~eV) of the surface. In addition, the spectrum exhibits a clear, but broad feature (A) centered at 0.6(1)~eV well above the secondary electron background. We approximate the latter with a biexponential background (dotted curve) and subtract it from the data. With the photon energy used in this experiment, feature A could originate (i) from an occupied initial state with a vertical binding energy of $E_{\mathrm{B}}= 2\cdot h\nu_{\mathrm{probe}}-E_{\mathrm{kin}}=6.6$~eV, (ii) from a normally unoccupied intermediate state $E_{\mathrm{B}}= h\nu_{\mathrm{probe}}-E_{\mathrm{kin}}=3.0$~eV, or (iii) a final state 0.6~eV above the vacuum level. In order to determine the nature of feature A, we first vary the duration of the laser pulses used in the experiment while keeping the photon fluence constant. This stretches the photon density in time.

If feature A were resulting from (i) two-photon photoemission from an initial state, its intensity should depend on the temporal photon density and become smaller upon increasing the laser pulse duration. The same behaviour would be expected if A would be a result of (ii) the transient population of an intermediate state with a short lifetime or (ii) a final state, as the photoemission signal would be generated by a two-photon process within one laser pulse in these cases, too. The inset of Fig.~\ref{lifetime}a shows the integrated intensity of peak A as a function of laser pulse duration (40 to 160 fs) at a constant number of photons per pulse. Within the accuracy of this experiment, the intensity of feature A remains constant. The absence of a dependence on the laser pulse duration proves that A is not (i) an occupied initial state, not (ii) a short-lived intermediate state A, and not (iii) a final state. In other words, A must be resulting from a long-lived intermediate state with a lifetime likely exceeding the inverse laser repetition rate. These observations are also consistent with the non-trivial fluence dependence of feature A, which will be subject of section~\ref{TTA}.

To test this hypothesis, we vary the repetition rate of the laser system. Fig.~\ref{lifetime}b depicts the spectral distribution of A for different laser repetition rates (200, 100 and 40 kHz), normalized by the fluence. Clearly, as the repetition rate decreases, the intensity of peak A drops and, in agreement with previous optical experiments\cite{Foglia2016}, a finite intensity remains for a time delay $\Delta T_{\mathrm{0}}=25~\mu \mathrm{s}=(40~\mathrm{kHz})^{-1}$. The inset of Fig.~\ref{lifetime}b shows the intensity evolution of A as a function of inverse laser repetition rate $\Delta T_{\mathrm{0}}$. The data can be fitted well by a single exponential decay function (dotted curve) yielding a characteristic time of $\tau_{\mathrm{C}}=15(4)~\mu$s. We conclude that peak A results from a normally unoccupied, transiently populated state with a lifetime exceeding the inverse laser repetition rate and likely coinciding with the DS probed in optical spectroscopy previously\cite{Foglia2016}. Therefore, upon photoexcitation, the SP6 film is in a \emph{photostationary} state, in which population and depopulation of A are balanced. Note that $\tau_{\mathrm{C}}$ does not necessarily reflect the lifetime of the excited state, as the variation of the repetition rate may alter the photostationary state non-trivially such that its intensity as a function of $\Delta T_0$ does not have to be dominated by the excited state lifetime.

\begin{figure}
\includegraphics[width=0.5\columnwidth]{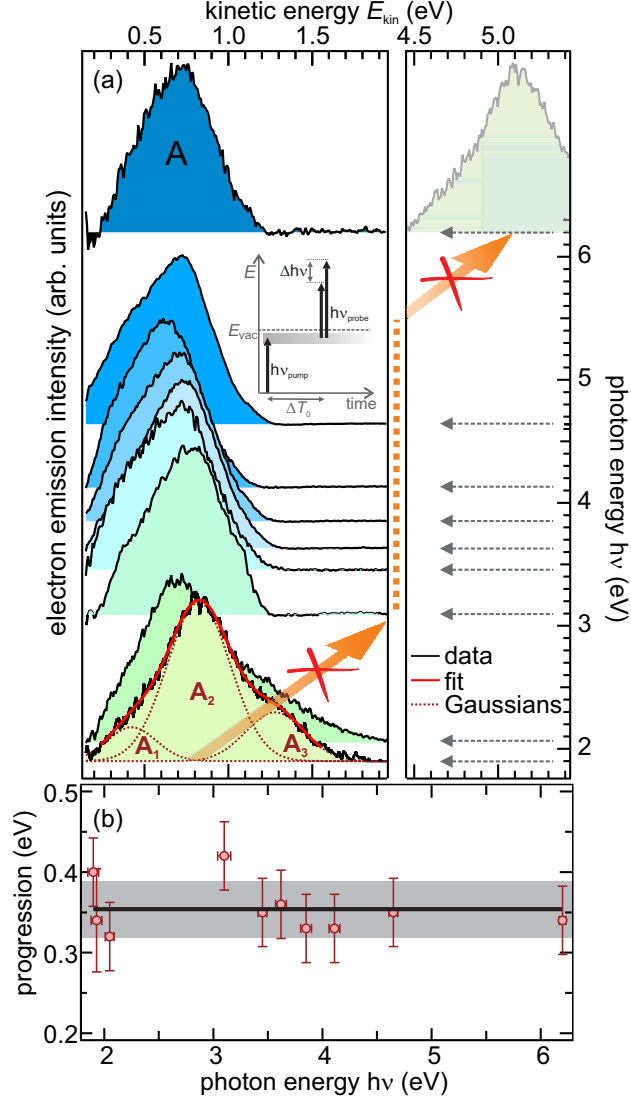}%
\caption{(a) Spectral position of the feature A for varying photon energies (1.9 - 6.2~eV). The expected spectral shift for a photoemission process (see illustration in the inset) is highlighted by the orange arrows. Spectra are shifted vertically for clarity and according to the photon energy used in the respective experiment (right axis). Feature A can be fit with the sum of three Gaussian peaks (red curves) separated by 360~meV as shown in panel (b).\label{hnu}}%
\end{figure}

When probed in photoemission, the spectral position of intermediate electronic states, i.e. the kinetic energy of the photoelectrons, depends on the photon energy h$\nu_{\mathrm{probe}}$ used to ionize the sample as illustrated by the diagram in Fig.~\ref{hnu}a. In single-color 2PPE, h$\nu_{\mathrm{probe}}$ is equal to h$\nu_{\mathrm{pump}}$ and a variation of the photon energy by $\Delta$h$\nu$ causes a shift of the spectral signature by the same amount. Fig.~\ref{hnu}a shows the spectral distribution of the feature A for different photon energies (right axis), ranging from 1.9 to 6.2~eV. Accordingly, if probed in a photoemission process, the spectral signature of the transiently populated, intermediate state A, measured using h$\nu_{\mathrm{probe}}=$1.9~eV (light green), should shift by $\Delta$h$\nu=4.3$~eV as indicated by the orange arrows. Clearly, this is not the case. Instead, peak A remains at $E_{\mathrm{kin}}=0.7(1)$~eV, corresponding to $E-E_{\mathrm{F}}=4.0(1)$~eV, where the slightly scattered position of the peak maximum is likely resulting from uncertainties in the background subtraction. \emph{This observation demonstrates that A is not probed in a photoemission process were electrons are emitted as a consequence of photon absorption}. Instead, the energy information of the laser pulse is “filtered” within the molecular system, making the emitted electrons expose the same $E_{\mathrm{kin}}$ irrespective of the photon energy used. This "filtering" must occur via an energy transfer process such as F\"orster or Dexter transfer, which are initiated by the photoexcitation and eventually lead to autoionization of SP6 molecules by the interaction of two excited states. 

The spectral shape of feature A can be fitted using the sum of three Gaussian peaks as exemplified for h$\nu_{\mathrm{probe}}=$~1.9~eV in Fig.~\ref{hnu}a. Likely, this progression is a result of a strong coupling of the involved electronic transitions to molecular vibrations in the SP6 molecule.  Fig.~\ref{hnu}b shows the energy difference of the two low-energy peaks, A$_1$ and A$_2$ as a function of photon energy, which is constant at 360(40)~meV. As shown previously\cite{Foglia2016}, the ground state vibrational progression of SP6 probed by photoluminescence spectroscopy is characterized by energy quanta of $\hbar \omega=$~170(10) and 180(10)~meV, which can be associated to CC-stretch and C-H-bend modes of the two central fluorenes in SP6 that are connected by the spiro-link (cf. Fig.~\ref{2PPE})\cite{Staehler2013}. We conclude that, apparently, two vibrational modes $\hbar \omega$ must be involved in the the autoionization process, likely distributed between two excited molecules. Remarkably, the high-energy peak A$_3$ only appears for photoexcitation using photons in the visible range, i.e. smaller photon energies than the absorption gap of SP6\cite{Blumstengel2008b}. In these experiments, the photoexcitation of SP6 requires the absorption of two photons and a significant electron emission signal thus $10^3$ times higher photon fluxes. At these high excitation densities also excited state absorption of A occurs (in a linear absorption process), thereby depositing a large amount of energy (heat) in the molecular film, which is reflected in the higher vibrational progression at these photon energies.

It should be noted that the comparably clear spectral resolution of the vibrational progression in feature A itself indicates that the electron emission is \emph{not} a photoemission process, but results from intermolecular energy transfer. As discussed in section~\ref{Ultrafast}, the amorphous structure of the SP6 film leads to an inhomogeneous broadening of the energy levels, which is reflected significantly more strongly in the photoelectron spectra that probe absolute energies than in the optical data which are sensitive to optical resonances. Similar to the latter, an intermolecular energy transfer is governed by energy differences, allowing to resolve the vibrational progression in the electron emission signal.

Autoionization by either a F\"orster or a Dexter energy transfer mechanism requires two interacting excited states. SP6 offers three potential candidates: X$_{\mathrm{6P}}$, X$_{\mathrm{2P}}$, exhibiting population dynamics on femto- and picosecond timescales and the long-lived dark state with a $\mu$s lifetime. In order to identify the excited states involved in the autoionization process, we performed autocorrelation experiments, where the time delay $\Delta t$ between two laser pulses of the same photon energy h$\nu =3.85$~eV was varied from fs to ns. The inset of Fig.~\ref{fluence} displays three exemplary spectra of the feature A: at $t=0(20)$~fs when pump and probe pulse overlap, at $\Delta t =150$~ps, and at $\Delta t =2$~ns. Clearly, the intensity of peak A is unaltered over this large range of timescales. As the ultrafast photoemission and optical experiments clearly showed (cf. section~\ref{Ultrafast} and Ref.~\cite{Foglia2016}) that both, X$_{\mathrm{6P}}$ and X$_{\mathrm{2P}}$ exhibit dramatic population changes on femto- and picosecond timescales, the sustained intensity of peak A up to 2~ns excludes any involvement of these states in the autoionization process. Thus, peak A can only be originating from long-lived species such as triplets (T), polarons (P), or charge-transfer excitons (CTX). While, in our experiments, we cannot unambiguously exclude CTX-CTX or even mixed interactions (CTX-T, T-P, CTX-P), autoionization of SP6 by T-T annihilation seems the most simple and likely process, a hypothesis which will be supported further below. Thus, for simplicity, we assume in the following analysis that the dark state is the SP6 triplet state. 

\subsection{TTA mediated by Dexter transfer}
\label{TTA}

The concentration of excited states does not change in a photostaionary state. Approximating the pulsed excitation at 200~kHz~$>(T_{\mathrm{0}})^{-1}$as continuous wave, the time-dependent change of the triplet concentration $n_{\mathrm{T}}$ must vanish, as described in literature previously\cite{Mikhnenko2015}:

\begin{equation}
    \frac{\mathrm{d}n_{\mathrm{T}}}{\mathrm{d} t} =\alpha n_{\mathrm{ph}}-k_{\mathrm{D}}n_{\mathrm{T}}-k_{\mathrm{TTA}}n_{\mathrm{T}}^2=0,
    \label{eq:DGL}
\end{equation}

\noindent In this case, all decay channels of $n_{\mathrm{T}}$ are balanced by the continuous repopulation. Here, $\alpha$ is the  population probability, $n_{\mathrm{ph}}$ the photon volume density, $k_{\mathrm{D}}$ the intrinsic triplet decay rate (excluding decay by TTA), and $k_{\mathrm{TTA}}$ the rate constant of TTA (in units of m$^3$s$^{-1}$). With the characteristic photon volume density\footnote{The volume V is calculated using the laser spot size $A=5.5\cdot10^{-8}~\mathrm{m}^2$, the optical penetration depth $\lambda=25$~nm\cite{Blumstengel2008b} in SP6, and the SP6 layer thickness of 25 nm}

\begin{equation}
   n_{\mathrm{c}}=\frac{k_{\mathrm{D}}^2}{k_{\mathrm{TTA}}}
    \label{eq:nC}
\end{equation}

\noindent that defines the cross-over between intrinsic triplet decay and TTA-dominated decay dynamics\cite{Mikhnenko2015}, the solution of Eq.~\ref{eq:DGL} becomes

\begin{equation}
    n_{\mathrm{T}}(n_{\mathrm{ph}})=\frac{1}{2}\frac{n_{\mathrm{c}}}{k_{\mathrm{D}}}\left(\sqrt{1+4\alpha\frac{ n_{\mathrm{ph}}}{n_{\mathrm{c}}}}-1\right).
    \label{eq:nT}
\end{equation}

\noindent The photon volume density dependence of the electron emission signal 

\begin{equation}
\begin{split}
    I_{\mathrm{EE}}(n_{\mathrm{ph}})&=p\cdot k_{\mathrm{TTA}}\cdot n_{\mathrm{T}}^2(n_{\mathrm{ph}})\cdot V\\
    &=pV\left(\frac{1}{2}n_{\mathrm{c}}+\alpha n_{\mathrm{ph}}-\sqrt{\left(\frac{1}{2}n_{\mathrm{c}}\right)^2+\alpha  n_{\mathrm{ph}}\cdot n_{\mathrm{c}}}\right)
    \label{eq:IEE}
\end{split}
\end{equation}

\noindent depends quadratically on the triplet density  $n_{\mathrm{T}}(n_{\mathrm{ph}})$ in the experiment. The prefactor $p$ accounts for the finite detection probability of emitted electrons and the population probability 

\begin{equation}
    \alpha=\frac{A_{\mathrm{2P}}}{A_{\mathrm{2P}}+A_{\mathrm{6P}}}=0.27
    \label{eq:alpha}
\end{equation}

\noindent can be approximated\footnote{This approximation assumes, following the line of arguments in Ref.~\cite{Foglia2016}, that all molecules excited to $X_{\mathrm{2P}}$ undergo ISC to the triplet state.} by the intensity ratio of the $X_{\mathrm{2P}}$ and $X_{\mathrm{6P}}$ optical resonances from Ref.~\cite{Foglia2016}.

\begin{figure}
\includegraphics[width=0.5\columnwidth]{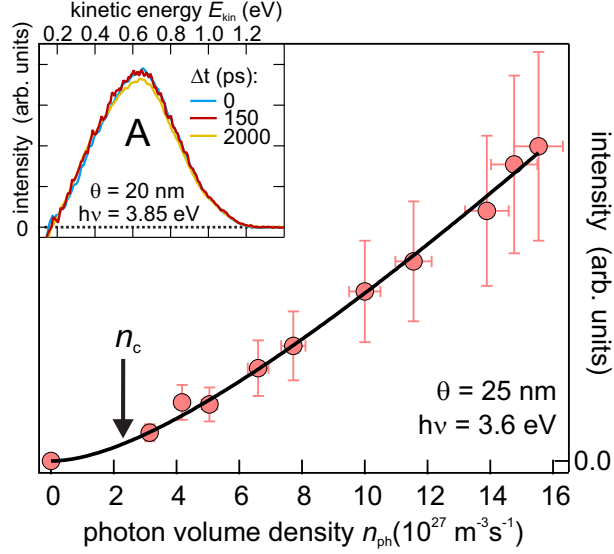}%
\caption{$n_{\mathrm{ph}}$ dependence of the electrons emitted by autoionization for a repetition rate of 200~kHz and a laser spot size of $5.5\cdot10^{-8}~\mathrm{m}^2$ (red markers) and the best fit to the data (black curve), see text for details. Inset: Autocorrelation experiment demonstrating the absence of dynamics on ultrafast timescales.  \label{fluence}}%
\end{figure}

Fig.~\ref{fluence} shows the dependence of the electron emission intensity on the absorbed photon volume density. If feature A were resulting from an ordinary one or two photon process within one laser pulse, its intensity should scale linearly or quadratically, respectively. This is not the case. When fitted with a power law (not shown), the electron emission intensity dependence on $n_{\mathrm{ph}}$ exposes an exponent of 1.5. This confirms the conclusion of the previous section that the detected spectral signature is not composed of \emph{photo}electrons, but that electron emission occurs as a consequence of excited state interaction. Using Eq.~\ref{eq:IEE}, where $n_{\mathrm{ph}}$, $V$ and $\alpha$ are known experimental parameters, the curvature of the $n_{\mathrm{ph}}$ dependence of the electron emission signal is solely determined by the characteristic photon volume density $n_{\mathrm{c}}$. Our least square fit (black) yields $n_{\mathrm{c}}=2.3(7)~10^{27}~\mathrm{m}^{-3}\mathrm{s}^{-1}$ as indicated by the arrow in Fig.~\ref{fluence}, showing that - for the photon densities applied - TTA is dominating the triplet decay. This is consistent with the major electron emission observed in the experiment.

In order to determine whether the TTA is mediated by F\"orster or Dexter transfer processes, we first calculate the quantum efficiency per triplet for TTA from our data using

\begin{equation}
\begin{split}
   \Phi_{\mathrm{TTA}}&=\frac{k_{\mathrm{TTA}}\cdot n_{\mathrm{T}}}{k_{\mathrm{D}}+k_{\mathrm{TTA}}\cdot n_{\mathrm{T}}}\\
   &=\frac{I_{\mathrm{EE}}(n_{\mathrm{ph}})}{\alpha n_{\mathrm{ph}}\cdot p V},
    \label{eq:p_tta}
\end{split}
\end{equation}

\begin{figure}
\includegraphics[width=0.5\columnwidth]{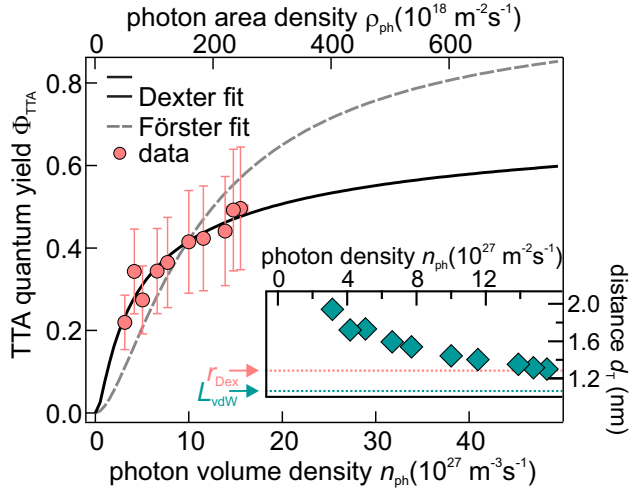}%
\caption{The photon density dependence of the experimental TTA quantum efficiency (red circles) is well-fitted for Dexter-type energy transfer \emph{en route} to autoionization (black curve). F\"orster-type transfer (dashed curve) cannot account for the slope of the experimental data. Inset: Average triplet distance for varying $n_{\mathrm{ph}}$.\label{dexter}}%
\end{figure}

\noindent which is a direct consequence of Eqs.~\ref{eq:DGL}-\ref{eq:IEE}, and plot it as a function of photon volume density (bottom axis) in Fig.~\ref{dexter} (red circles). Remarkably, the TTA quantum efficiency lies between 22 and 50~$\%$ for the excitation densities used in the experiment. For comparison, we also provide the corresponding photon area densities $\rho_{\mathrm{ph}}$ in the top axis. 

With Eqs.~\ref{eq:kF}-\ref{eq:QY} for F\"orster and Dexter energy transfer we can also formulate mathematical expressions for the TTA quantum efficiency in the case of Dexter energy transfer

\begin{equation}
   \Phi_{\mathrm{TTA}}^{\mathrm{Dex}}(n_{\mathrm{ph}})=\left[1+\mathrm{exp}\left(-\frac{2r_{\mathrm{Dex}}}{L}\left(1-\frac{d_{\mathrm{T}}(n_{\mathrm{ph}})}{r_{\mathrm{Dex}}}\right)\right)\right]^{-1}
    \label{eq:pTTADex}
\end{equation}

\noindent and in the case of Förster energy transfer

\begin{equation}
   \Phi_{\mathrm{TTA}}^{\mathrm{F}}(n_{\mathrm{ph}})=\left[1+\left(\frac{d_{\mathrm{T}}(n_{\mathrm{ph}})}{r_{\mathrm{F}}}\right)^6\right]^{-1},
    \label{eq:pTTAF}
\end{equation}

\noindent where

\begin{equation}
   d_{\mathrm{T}}(n_{\mathrm{ph}})=2\cdot\sqrt[3]{\frac{3}{4\pi\cdot n_{\mathrm{T}}(n_{\mathrm{ph}})}}.
    \label{eq:rT}
\end{equation}

\noindent These can be used to fit the experimental $\Phi_{\mathrm{TTA}}$ in Fig.~\ref{dexter} (solid and dashed curve, respectively). We approximate the mean van-der-Waals distance between two SP6 molecules by

\begin{equation}
   L=2\cdot\sqrt[3]{\frac{3}{4\pi\cdot \rho}}
    \label{eq:vdW}
\end{equation}

\noindent with the SP6 density $\rho=1.14\cdot 10^{27}\mathrm{m}^{-3}$. Clearly, the distance dependence of the F\"orster-type transfer (grey dashed curve) does not follow the experimental trend and can, therefore, be excluded. The quantum efficiency for Dexter-mediated TTA (Eq.~\ref{eq:pTTADex}, black solid curve), on the contrary, provides a perfect fit to the data with only two free parameters. It yields a Dexter radius of $r_{\mathrm{Dex}}=$~1.3(2)~nm and an intrinsic triplet lifetime of $\tau_{\mathrm{D}}=k_{\mathrm{D}}^{-1}=0.10(3)$~s. This enables the calculation of the rate constant for TTA based on Eq.~\ref{eq:nC} to be only $k_{\mathrm{TTA}}=4.5~10^{-26}~\mathrm{m}^3\mathrm{s}^{-1}$. This extraordinarily small value lies orders of magnitude below typical literature values. The nevertheless significant TTA quantum efficiency (Fig.~\ref{dexter}) can only be resulting from an extraordinarily high triplet density.

With Eqs.~\ref{eq:rT} and \ref{eq:nT}, we can calculate the average triplet distance as a function of photon density as displayed in the inset of Fig.~\ref{dexter}. For all excitation densities used in our experiment the triplet-triplet distance is with less than 2~nm very small. With increasing fluence, it even decreases and approximates the Dexter radius $r_{\mathrm{Dex}}$ (red) and the van-der-Waals distance of two SP6 molecules $L$. These small distances, i.e. high triplet densities explain the with 50$\%$ large TTA quantum efficiency despite the extremely low TTA rate constant $k_{\mathrm{TTA}}$.

\begin{figure}
\includegraphics[width=0.5\columnwidth]{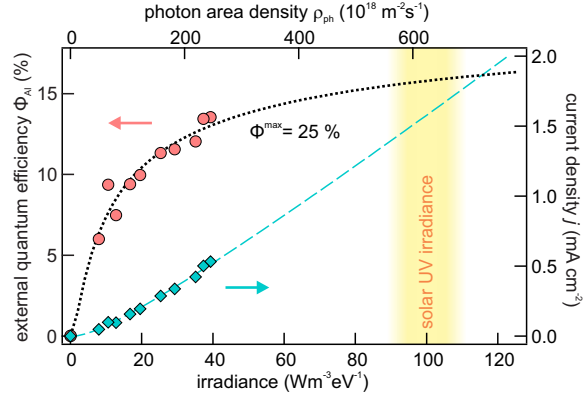}%
\caption{EQE (left axis) and current density (right axis) for illumination of condensed SP6 as measured (markers) using h$\nu=3.6$~eV and calculated based on the TTA model with Dexter-type energy transfer (dotted and dashed curves). \label{QuY}}%
\end{figure}

Based on the above, it is possible to calculate the external quantum efficiency (EQE) for photoinduced electron emission through autoionization

\begin{equation}
   \Phi_{\mathrm{AI}}=\alpha\cdot \Phi_{\mathrm{TTA}},
    \label{eq:QY_AI}
\end{equation}

\noindent which is plotted (circles) in Fig.~\ref{QuY} as a function of incident photon area density (top axis) and irradiance (bottom axis). We calculate the expected excitation density-dependent evolution of the EQE for the autoionization process based on the above model using Eqs.~\ref{eq:pTTADex} and \ref{eq:QY_AI}. This is displayed by the dotted curve in Fig.~\ref{QuY} and extrapolates to a maximum of 25~$\%$ for highest photon fluxes. Fig.~\ref{QuY} also shows the experimental (diamonds) and extrapolated current densities $j=\Phi_{\mathrm{AI}}\cdot\rho_{\mathrm{ph}}\cdot e$ (right axis) where $e$ is the elementary charge. The highest current density in our experiments amounts to 0.5~mA~cm$^{-2}$. In the range of solar irradiance on the order of 100~W~$\mathrm{m}^{-2}~\mathrm{eV}$ at similar photon energies\cite{GUEYMARD2004}, the extrapolated EQE for electron emission and the corresponding current density $j$ are on the order of $15~\%$ and 1.5~mA~cm$^{-2}$, respectively.

\begin{table}
\caption{Summary of parameters extracted using the TTA model\label{table} }
\begin{tabular}{lccr}
\toprule
IVR time (fast) & $\tau_{\mathrm{IVR1}}$ & $200-400$~fs\\
IVR time (slow) & $\tau_{\mathrm{IVR2}}$ & $2-6$~ps\\
ISC time & $\tau_{\mathrm{ISC}}$ & $240(30)$~ps\\
triplet lifetime & $\tau_{\mathrm{D}}$ & $0.10(3)$~s\\
TTA rate constant & $k_{\mathrm{TTA}}$ & $4.5\cdot 10^{-26}\mathrm{m}^{3}\mathrm{s}^{-1}$\\
Dexter radius & $r_{\mathrm{Dex}}$ & 1.3(2)~nm\\
EQE...\\
...at solar UV irradiances & $\Phi_{\mathrm{AI}}^{\mathrm{solar}}$ & $15~\%$\\
...maximum & $\Phi_{\mathrm{AI}}^{\mathrm{max}}$ & $25~\%$\\
current density \\
...at solar UV irradiances & $j_{\mathrm{solar}}$ & 1.5~mA~cm$^{-2}$\\

\end{tabular}
\end{table}

In summary, the TTA model describes the excitation density-dependent evolution of our data very well, allowing for the calculation of the TTA quantum efficiency as a function of photon density, which behaves as expected for Dexter-type energy transfer of the triplets. The very high probability for a triplet to undergo TTA ($49~\%$ at the largest triplet density in our experiments) can be translated to a sizeable EQE of $\Phi_{\mathrm{AI}}=\alpha\cdot \Phi_{\mathrm{TTA}}=$~13$~\%$ at highest experimental excitation densities, which corresponds to a current density $j=0.5$~mA~cm$^{-2}$.  

It should be noted that the above model is not restricted to TTA, but could be directly translated to CTX-CTX annihilation. For electron emission due to mixed interactions, such as CTX-P (cf. section~\ref{AI}), the mathematical framework would stay similar - albeit with a larger number of free parameters. The extracted high triplet density in our experiments, very close to the SP6 density, displayed in the inset of Fig.~\ref{dexter}, however, is a strong indication that the underlying mechanism is indeed TTA, as such a dense distribution of CTX with or without polarons, is physically not possible.

\subsection{Discussion and Conclusions}
\label{discussion}

\begin{figure}
\includegraphics[width=0.6\columnwidth]{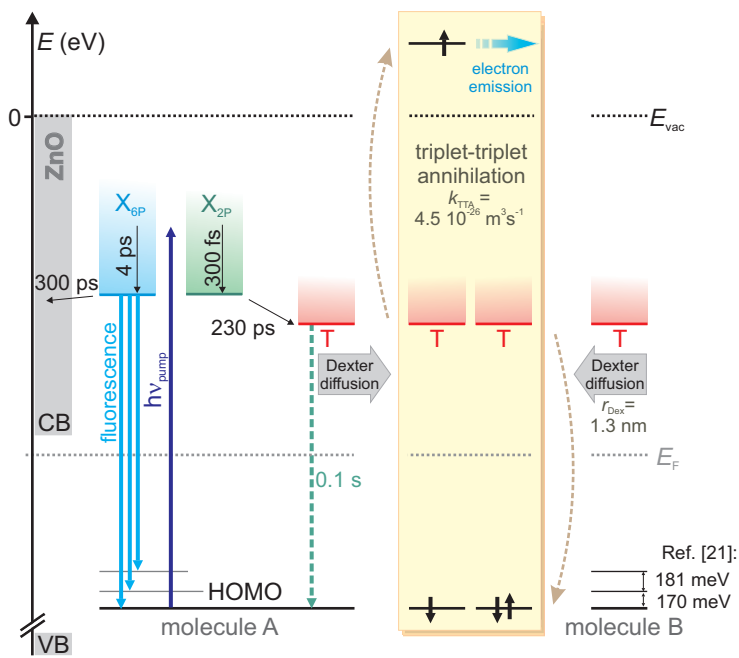}%
\caption{Overview of the elementary processes occuring in SP6 film upon UV photoexcitation. See text for details.\label{summary}}%
\end{figure}

Fig.~\ref{summary} summarizes the pathways of optical excitations in SP6 layers and table~\ref{table} displays all relevant experimental parameters determined in this work. As shown previously\cite{Foglia2016}, photoexcitation of SP6 leads to the population of two separate electronic states, X$_{\mathrm{6P}}$ and X$_{\mathrm{2P}}$. Excess energy is released by internal vibrational relaxation on femto- and picosecond timescales, which is equivalently monitored by time-resolved photoelectron and optical spectroscopy (cf. section~\ref{Ultrafast}). X$_{\mathrm{6P}}$, which is localized on the 6P unit of the molecule, decays by fluorescence and diffusion-limited electron transfer to the ZnO substrate. X$_{\mathrm{2P}}$ solely decays into a dark state with a time constant of $\tau_{\mathrm{ISC}}=$~240(30)~ps that is consistently measured by photoelectron and optical spectroscopy. This dark state, likely the SP6 triplet, exhibits a long lifetime, which enables Dexter-type energy transfer of the triplet excitation. Beyond relaxation, it decays by TTA leading to autoionization of SP6 molecules. The resulting electron emission appears as a photon energy-independent peak in the photoelectron spectra and exhibits a vibrational progression with energy quanta of twice the CC-stretch and CH-bend modes dominating the vibrational ground state progression\cite{Staehler2013}.

Based on excitation density-dependent experiments and a simple rate equation model, we determine the characteristic photon density for TTA, the intrinsic triplet lifetime, TTA rate constant, Dexter radius, as well as the EQE and current density in the experiment as listed in table~\ref{table}. The latter two are, with 15$~\%$ and 1.5~mA~cm$^{-2}$ at solar UV irradiances, respectively, quite high for an illuminated, pure and mono-molecular organic film, which is not a complex heterostructure as in current organic light harvesting devices. As the chemical potential  between the electron and hole after successful TTA is larger than 4.0~eV and, thus, very high, the power conversion efficiency

\begin{equation}
   \mathrm{PCE} = \mathrm{FF}\cdot\frac{P_{\mathrm{out}}}{\mathrm{h}\nu\cdot \rho_{\mathrm{ph}}^{\mathrm{solar}}}
    \label{eq:PCE}
\end{equation}

\noindent where FF is the unknown fill factor and $P_{\mathrm{out}}= 4~\mathrm{V}\cdot j_{\mathrm{solar}}$ the output power achieved under illumination with a photon density of $\rho_{\mathrm{ph}}^{\mathrm{solar}} = 600\cdot 10^{18}~\mathrm{m}^{-2}\mathrm{s}^{-1}$ at $\mathrm{h}\nu=3.6$~eV, the power conversion efficiency at this UV photon energy becomes PCE~$\geq $~FF$\cdot 17\%$. This value is - estimated for typical fill factors clearly above 0.5 - a compatible value for organic photovoltaics that show PCEs up to 17~$\%$\cite{Meng2018}. Based on this, we conclude that SP6 is a promising material for forthcoming organic tandem cells. Moreover, this work is the proof of principle of highly efficient electron emission through TTA. We hope that our work stimulates the design of new organic molecules that exhibit or even exceed the fundamentally necessary properties of SP6 that enable efficient autoionization: matching binding energies $E_{\mathrm{B}}^{\mathrm{T}}<\frac{1}{2}\cdot E_{\mathrm{B}}^{\mathrm{S}_{0}}$ of triplet and ground state with respect to the ionization potential, respectively, characteristic photon densities $n_{\mathrm{c}}$ in the 10$^{27}$~m$^{-3}$~s$^{-1}$ range, and a large packing density of the molecules. Organic materials with these properties and possibly even absorption in the \emph{visible} light range seem highly promising for future light-harvesting applications.

\begin{acknowledgments}
Funded by the Deutsche Forschungsgemeinschaft (DFG, German Research Foundation) - Projektnummer 182087777 - SFB 951. Additional funding by the European Union through grant No. 280879-2 CRONOS CP-FP7. The authors thank Emil List-Kratochvil for fruitful discussions, Sylke Blumstengel for providing the SP6 molecules and Mino Sparenberg for their density determination, as well as Selene Mor for technical support during autocorrelation experiments. SV would like to thank Max Planck Research Society for a Max Planck Postdoctoral Fellowship.
\end{acknowledgments}


%
%

%


\bibliography{Vempati_etal_SP6.bib}

\end{document}